\documentclass[preprint,showpacs,preprintnumbers,amsmath,amssymb,nofootinbib]{revtex4-1}

\usepackage{graphicx}
\usepackage{dcolumn}
\usepackage{amssymb}
\usepackage{mathrsfs}
\usepackage{amsmath}
\usepackage{epsfig}
\usepackage{hhline}
\usepackage[utf8]{inputenc}
\usepackage{color}
\usepackage{multirow}
\usepackage[T1]{fontenc}
\usepackage{verbatim}
\usepackage{slashed}

\begin{document}

\title{Polarization of top quark in vector-like quark decay}

\author{Hang Zhou}
\author{Ning Liu}
\email{corresponding author: wlln@mail.ustc.edu.cn}
\affiliation{School of Physics Science and Technology, Nanjing Normal University, Nanjing, 210023, China}

\begin{abstract}
Vector-like quarks\,(VLQs) are attractive extensions to the Standard Model. They mix with the SM quarks and can lead to rich phenomenology. Determination of VLQ's interaction structure with the SM is then an important issue, which can be inferred from the decay products of VLQs, such as top quark. We calculate the spin-analyzing powers for charged leptons from top quark in VLQ decay for various VLQ scenarios. We find that the top polarization effect will be helpful to distinguish different natures of the VLQ couplings with the SM.
\end{abstract}
\maketitle

\section{Introduction}

Extra vector-like quarks (VLQ) are usually expected to be present in many beyond-SM models, such as little Higgs models, composite Higgs models and some extra-dimension models. For example, models are proposed to explain the lightness of the observed Higgs by assuming that it is a pseudo Goldstone boson and VLQs generally appear in these theories\,\cite{Perelstein:2003wd,Contino:2006qr,Matsedonskyi:2012ym}. In some supersymmetric models, the introduction of vector-like quarks can relax the restrictions imposed on the MSSM by the 125 GeV Higgs\,\cite{Moroi:1992zk,Babu:2008ge,Martin:2012dg}. As a popular extension to the SM, VLQs in a variety of models have been widely studied\,\cite{Cheng:2005as,Han:2003wu,Atre:2011ae,Atre:2008iu,Cao:2006wk,Cao:2007ea,Han:2016bus,Liu:2015kmo,Han:2014qia,Kearney:2013oia,Kearney:2013cca,Kats:2017ojr,Barducci:2017xtw}.

These new quarks are triplets under the $SU(3)_{C}$ gauge group just like the SM quarks, but have the same electroweak quantum numbers for both left- and right-handed components. That is, VLQs of different chiralities transform in the same representation of $SU(2)$. In a model-independent way, vector-like quarks can generally be introduced as $T$, $B$, $X$ and $Y$: $T$ and $B$ are quarks with electric charges of $+2/3$ and $-1/3$ respectively, while $X$ and $Y$ are ones of charge $+5/3$ and $-4/3$ respectively, which appear as different $SU(2)$ multiplets\,\cite{delAguila:2000aa,delAguila:2000rc}. VLQs' interaction in different multiplets have been studied systematically in \,\cite{AguilarSaavedra:2009es}. They are generally considered to mix with the SM quarks and thus can be involved in flavor-changing neutral currents (FCNC) at tree level\,\cite{delAguila:1982fs,Branco:1986my}, which receives strong constraints from experiments.

At the LHC and other colliders, extra new quarks have long been searched for. The chirally coupled new quarks have already been excluded by recent searches and tests\,\cite{Djouadi:2012ae,Eberhardt:2012gv}, whereas the vector-like quarks survived. At a proton-proton collider, VLQs can be produced singly or in pairs and then decay to a SM quark and a gauge boson or a Higgs boson\,\cite{AguilarSaavedra:2009es}. Recent searches for the $+2/3$ charged $T$ quark and $-4/3$ charged $Y$ quark from $13$ \,TeV pp collision data give a lower limit for their masses at $1.3$\,TeV at 95\%C.L.\,\cite{Sirunyan:2017pks}, while for the $-1/3$ charged $B$ quark, current analysis pushes the lower limit of $B$ mass up to $1.5$\,TeV from its single production at pp collision\,\cite{Tanabashi:2018oca,Aad:2015voa}. As for the $5/3$ charged $X$ quark, pair-production searches based on $13$\,TeV pp collision set the mass limit at $1.02$\,TeV\,\cite{Tanabashi:2018oca,Sirunyan:2017jin}. It should be noted that the VLQs once produced will decay into the SM quarks, which are generally polarized due to VLQs' parity-violating couplings with the gauge bosons. Different from other SM quarks, top quarks decay before hadronization and hence the top polarization can be measured by the kinematics of its decay products. As a result, the final states from top decay can in turn provide information about the parent VLQs' gauge interaction, like the cases that have been studied in top decay or top-squark decay\,\cite{V.:2016wba,Belanger:2012tm,Choudhury:2010cd,Godbole:2006tq,Han:2008gy,Barger:2006hm,Arai:2009cp,Arai:2007ts,Liu:2010zze,Gopalakrishna:2010xm,Bernreuther:2006pd,Bernreuther:2010ny,Bernreuther:2013aga,Bernreuther:2015yna,Bernreuther:2017yhg,He:2007tt,Cao:2015doa,Wu:2018xiz}. In this paper, we study, in a model-independent way, the polarization effects in the decay of VLQs in two scenarios:
\begin{align}
\textrm{Singlets}:{}&\qquad T_{L,R}\,,\quad B_{L,R}\,,\notag\\
\textrm{Doublets}:{}&\qquad (X\,T)_{L,R}\,,\quad (T\,B)_{L,R}\,,\quad (B\,Y)_{L,R}\,.
\end{align}
These polarization effects can be used to differentiate the $SU(2)$ nature of these new quarks if ever discovered. 

This paper is organized as follows. In section II we introduce the VLQ-related interactions in different scenarios mentioned above, which determine the decay modes of VLQs. Then we calculate the spin analyzing power of the final charged leptons from VLQ decays and analyze the polarization effects in different scenarios in section III. Section IV is our conclusion.

\section{Relevant interactions in different VLQ scenarios}

As stated above, the vector-like quarks are studied in this paper in two different multiplets: singlet and doublet. We give in this section the relevant Lagrangian in terms of mass eigenstates of quarks with gauge bosons and the Higgs boson. 

\subsection{Singlets: $T$ singlet and $B$ singlet}

\subsubsection{$T$ singlet}
In this section and the rest of the paper, Greek indices $\alpha,\beta=1,2,3,4$ run over all quarks including the new vector-like quarks, while Latin indices $i,j=1,2,3$ over three generations of the SM quarks. An introduction of a singlet VLQ $T$ leads to the $W$-$T$, $Z$-$T$ and Higgs-$T$ couplings as follows,
\begin{align}
\mathcal{L}_{W-\textrm{quark}}={}&-\frac{g}{\sqrt{2}}\bar{u}_{L\alpha}\gamma^{\mu}V_{\alpha i}d_{Li}W^{+}_{\mu}+\textrm{H.c.}\,,\label{T-W}\\
\mathcal{L}_{Z-\textrm{quark}}={}&-\frac{g}{2\cos\theta_{W}}\left(\bar{u}_{L\alpha}\gamma^{\mu}X_{\alpha\beta}u_{L\beta}-2s^{2}_{W}J^{\mu}_{\textrm{EM}}\right)Z_{\mu}\,,\label{T-Z}\\
\mathcal{L}_{\textrm{H}-\textrm{quark}}={}&-\frac{g}{2m_{W}}\left(m_{u,\beta}\bar{u}_{L\alpha}X_{\alpha\beta}u_{R\beta}+m_{u,\alpha}\bar{u}_{R\alpha}X_{\alpha\beta}u_{L\beta}\right)H\,,\label{T-H}
\end{align}
in which $J^{\mu}_{\textrm{EM}}$ is the electromagnetic current and sums over all quarks. The CKM quark-mixing matrix $V$ is generalized to dimension $4\times3$ to include the new vector-like ones and $X=VV^{\dag}$ is a $4\times4$ Hermitian matrix. $m_{u,\alpha}$ is mass of the up-type quark. $\theta_{W}$ is the Weinberg angle. In this $T$ singlet scenario, interactions given above lead to decays of T quark into $W^{+}b$, $Zt$ and $Ht$. From \eqref{T-W}-\eqref{T-H}, we can write explicitly the terms determining these decay processes
\begin{align}
\mathcal{L}_{T\rightarrow Wb}={}&-\frac{g}{\sqrt{2}}\bar{T}\gamma^{\mu}V_{Tb}P_{L}bW^{+}_{\mu}+\textrm{H.c.}\,,\\
\mathcal{L}_{T\rightarrow Zt}={}&-\frac{g}{2\cos\theta_{W}}\bar{t}\gamma^{\mu}X_{tT}P_{L}TZ_{\mu}\,,\\
\mathcal{L}_{T\rightarrow Ht}={}&-\frac{gX_{tT}}{2m_{W}}\bar{t}\left(m_{T}P_{R}+m_{t}P_{L}\right)TH\,.
\end{align}

\subsubsection{$B$ singlet}

The coupling terms in the $B$ singlet scenario are similar to ones in the above $T$ singlet scenario.
\begin{align}
\mathcal{L}_{W-\textrm{quark}}={}&-\frac{g}{\sqrt{2}}\bar{u}_{Lj}\gamma^{\mu}V_{j\alpha}d_{L\alpha}W^{+}_{\mu}+\textrm{H.c.}\,,\\
\mathcal{L}_{Z-\textrm{quark}}={}&\frac{g}{2\cos\theta_{W}}\left(\bar{d}_{L\alpha}\gamma^{\mu}X_{\alpha\beta}d_{L\beta}+2s^{2}_{W}J^{\mu}_{\textrm{EM}}\right)Z_{\mu}\,,\\
\mathcal{L}_{\textrm{H}-\textrm{quark}}={}&-\frac{g}{2m_{W}}\left(m_{d,\beta}\bar{d}_{L\alpha}X_{\alpha\beta}d_{R\beta}+m_{d,\alpha}\bar{d}_{R\alpha}X_{\alpha\beta}d_{L\beta}\right)H\,,
\end{align}
In this scenario the CKM matrix $V$ is a $3\times4$ matrix and $X=V^{\dag}V$. $m_{d,\alpha}$ is mass of the down-type quark. Interactions in this scenario lead to decays of $B$ quark into $W^{-}t$, $Zb$ and $Hb$, the Lagrangian of which can be specifically expressed as follows
\begin{align}
\mathcal{L}_{B\rightarrow Wt}={}&-\frac{g}{\sqrt{2}}\bar{t}\gamma^{\mu}V_{tB}P_{L}BW^{+}_{\mu}+\textrm{H.c.}\,,\\
\mathcal{L}_{B\rightarrow Zb}={}&\frac{g}{2\cos\theta_{W}}\bar{b}\gamma^{\mu}X_{bB}P_{L}BZ_{\mu}\,,\\
\mathcal{L}_{B\rightarrow Hb}={}&-\frac{gX_{bB}}{2m_{W}}\bar{b}\left(m_{B}P_{R}+m_{b}P_{L}\right)BH\,.
\end{align}

\subsection{Doublets: $(T B)$, $(X T)$ and $(Y B)$ doublet}

\subsubsection{$(T B)$ doublet}
In the $(T B)$ doublet scenario, the relevant couplings written in mass eigenstates are
\begin{align}
\mathcal{L}_{W-\textrm{quark}}={}&-\frac{g}{\sqrt{2}}\left(\bar{u}_{Li}\gamma^{\mu}V_{L,ij}d_{Lj}+\bar{u}_{R\alpha}\gamma^{\mu}V_{R,\alpha\beta}d_{R\beta}+\bar{T}_{L}\gamma^{\mu}B_{L}\right)W^{+}_{\mu}+\textrm{H.c.}\,,\\
\mathcal{L}_{Z-\textrm{quark}}={}&-\frac{g}{2\cos\theta_{W}}\big(\bar{u}_{L\alpha}\gamma^{\mu}u_{L\alpha}-\bar{d}_{L\alpha}\gamma^{\mu}d_{L\alpha}\notag\\
{}&+\bar{u}_{R\alpha}\gamma^{\mu}X_{u,\alpha\beta}u_{R\beta}-\bar{d}_{R\alpha}\gamma^{\mu}X_{d,\alpha\beta}d_{R\beta}-2s^{2}_{W}J^{\mu}_{\textrm{EM}}\big)Z_{\mu}\,,\\
\mathcal{L}_{\textrm{H}-\textrm{quark}}={}&-\frac{g}{2m_{W}}\big[m_{u,\alpha}\bar{u}_{L\alpha}(\delta_{\alpha\beta}-X_{u,\alpha\beta})u_{R\beta}+m_{u,\beta}\bar{u}_{R\alpha}(\delta_{\alpha\beta}-X_{u,\alpha\beta})u_{L\beta}\notag\\
{}&+m_{d,\alpha}\bar{d}_{L\alpha}(\delta_{\alpha\beta}-X_{d,\alpha\beta})d_{R\beta}+m_{d,\beta}\bar{d}_{R\alpha}(\delta_{\alpha\beta}-X_{d,\alpha\beta})d_{L\beta}\big]H\,,
\end{align}
where $V_{L}$ is the $3\times3$ CKM quark-mixing matrix and $V_{R}$ is a $4\times4$ generalized mixing matrix. $X_{u}=V_{R}V^{\dag}_{R}$ and $X_{d}=V^{\dag}_{R}V_{R}$ are Hermitian and non-diagonal leading to FCNC. The above interactions lead to $T$ decays into $W^{+}b$, $Zt$ and $Ht$, while $B$ decays into $W^{-}t$, $Zb$ and $Hb$. These interactions can be expressed specifically
\begin{align}
\mathcal{L}_{T\rightarrow Wb,B\rightarrow Wt}={}&-\frac{g}{\sqrt{2}}\left(\bar{t}\gamma^{\mu}V_{R,tB}P_{R}B+\bar{T}\gamma^{\mu}V_{R,Tb}P_{R}b\right)W^{+}_{\mu}+\textrm{H.c.}\,,\\
\mathcal{L}_{T\rightarrow Zt,B\rightarrow Zb}={}&\frac{g}{2\cos\theta_{W}}\left(\bar{b}\gamma^{\mu}X_{d,bB}P_{R}B-\bar{t}\gamma^{\mu}X_{u,tT}P_{R}T\right)Z_{\mu}\,,\\
\mathcal{L}_{T\rightarrow Ht,B\rightarrow Hb}={}&\frac{g}{2m_{W}}\left[\bar{t}X_{u,tT}\left(m_{t}P_{R}+m_{T}P_{L}\right)T+\bar{b}X_{d,bB}\left(m_{b}P_{R}+m_{B}P_{L}\right)B\right]H\,.
\end{align}

\subsubsection{$(X T)$ doublet}
In this case, $+5/3$ charged quark $X$ and $+2/3$ charged quark $T$ form a $SU(2)$ doublet with a hypercharge $7/6$. The relevant terms of Lagrangian are
\begin{align}
\mathcal{L}_{W-\textrm{quark}}={}&-\frac{g}{\sqrt{2}}\left(\bar{u}_{Li}\gamma^{\mu}V_{L,ij}d_{Lj}+\bar{X}_{L}\gamma^{\mu}T_{L}+\bar{X}_{R}\gamma^{\mu}V_{R,X\beta}u_{R\beta}\right)W^{+}_{\mu}+\textrm{H.c.}\,,\\
\mathcal{L}_{Z-\textrm{quark}}={}&-\frac{g}{2\cos\theta_{W}}\left(\bar{u}_{Li}\gamma^{\mu}u_{Li}-\bar{u}_{R\alpha}\gamma^{\mu}X_{\alpha\beta}u_{R\beta}-\bar{T}_{L}\gamma^{\mu}T_{L}+\bar{X}\gamma^{\mu}X-2s^{2}_{W}J^{\mu}_{\textrm{EM}}\right)Z_{\mu}\,,\\
\mathcal{L}_{\textrm{H}-\textrm{quark}}={}&-\frac{g}{2m_{W}}\left[m_{u,\alpha}\bar{u}_{L\alpha}(\delta_{\alpha\beta}-X_{\alpha\beta})u_{R\beta}+m_{u,\beta}\bar{u}_{R\alpha}(\delta_{\alpha\beta}-X_{\alpha\beta})u_{L\beta}\right]H\,,
\end{align}
where $V_{R}$ is a $1\times4$ matrix and $X=V^{\dag}_{R}V_{R}$. In the $(X T)$ doublet scenario, $T$ quark decays into $Zt$ and $Ht$, while $X$ quark decays into $W^{+}t$, considering an almost degenerate mass spectrum $m_{X}\sim m_{T}$. Coupling terms relevant to these processes are
\begin{align}
\mathcal{L}_{X\rightarrow Wt}={}&-\frac{g}{\sqrt{2}}\bar{t}\gamma^{\mu}V^{*}_{R,Xt}P_{R}XW^{-}_{\mu}+\textrm{H.c.}\,,\\
\mathcal{L}_{T\rightarrow Zt}={}&\frac{gX_{tT}}{2\cos\theta_{W}}\bar{t}\gamma^{\mu}P_{R}TZ_{\mu}\,,\\
\mathcal{L}_{T\rightarrow Ht}={}&\frac{gX_{tT}}{2m_{W}}\bar{t}\left(m_{t}P_{R}+m_{T}P_{L}\right)TH\,.
\end{align}

\subsubsection{$(B Y)$ doublet}
Similarly extra $-1/3$ charged quark $B$ and $-4/3$ charged quark $Y$ can form a $SU(2)$ doublet with a hypercharge $-5/6$, we can write down in this case the relevant couplings
\begin{align}
\mathcal{L}_{W-\textrm{quark}}={}&-\frac{g}{\sqrt{2}}\left(\bar{u}_{Li}\gamma^{\mu}V_{L,ij}d_{Lj}+\bar{B}_{L}\gamma^{\mu}Y_{L}+\bar{d}_{R\alpha}\gamma^{\mu}V_{R,\alpha Y}Y_{R}\right)W^{+}_{\mu}+\textrm{H.c.}\,,\\
\mathcal{L}_{Z-\textrm{quark}}={}&-\frac{g}{2\cos\theta_{W}}\left(-\bar{d}_{Lj}\gamma^{\mu}d_{Lj}+\bar{d}_{R\alpha}\gamma^{\mu}X_{\alpha\beta}d_{R\beta}+\bar{B}_{L}\gamma^{\mu}B_{L}-\bar{Y}\gamma^{\mu}Y-2s^{2}_{W}J^{\mu}_{\textrm{EM}}\right)Z_{\mu}\,,\\
\mathcal{L}_{\textrm{H}-\textrm{quark}}={}&-\frac{g}{2m_{W}}\left[m_{d,\alpha}\bar{d}_{L\alpha}(\delta_{\alpha\beta}-X_{\alpha\beta})d_{R\beta}+m_{d,\beta}\bar{d}_{R\alpha}(\delta_{\alpha\beta}-X_{\alpha\beta})d_{L\beta}\right]H\,,
\end{align}
in which $V_{R}$ is a $4\times1$ matrix and $X=V_{R}V^{\dag}_{R}$. Allowed decays are $Y\rightarrow W^{-}b$ and $B\rightarrow Zb, Hb$ considering an almost degenerate mass spectrum $m_{B}\sim m_{Y}$. Coupling terms relevant to these processes are
\begin{align}
\mathcal{L}_{Y\rightarrow Wb}={}&-\frac{g}{\sqrt{2}}\bar{b}\gamma^{\mu}V_{R,bY}P_{R}YW^{+}_{\mu}+\textrm{H.c.}\,,\\
\mathcal{L}_{B\rightarrow Zb}={}&-\frac{gX_{bB}}{2\cos\theta_{W}}\bar{b}\gamma^{\mu}P_{R}BZ_{\mu}\,,\\
\mathcal{L}_{B\rightarrow Hb}={}&\frac{gX_{bB}}{2m_{W}}\bar{b}\left(m_{b}P_{R}+m_{B}P_{L}\right)BH\,.
\end{align}

\section{Top polarization in VLQ decays and spin-analyzing power of the charged lepton}

As we have mentioned, top quarks from VLQ decays are polarized due to the parity-violating couplings and this polarization effect can be measured by the decay products of top quark due to its short lifetime. We calculate in different VLQ scenarios the spin-analyzing power for the final charged lepton in the decay chains of VLQ. The spin-analyzing power is generally defined as follows: the angular distribution of a decay product $f$ in the parent rest frame is given by,
\begin{eqnarray}
\frac{1}{\Gamma}\frac{d\Gamma}{d\cos\theta_{f}}=\frac{1}{2}(1+ \mathcal{P}_{f}\cos\theta_{f}),
\end{eqnarray}
in which $\theta_{f}$ is the angle between the momentum of particle $f$ in the final state and the spin vector of the decaying particle in its rest frame, the coefficient $\mathcal{P}_{f}$ is the spin analyzing power of final-state particle $f$. In the following, we compute the angular distributions of the charged lepton from the decays of VLQs. An on-shell top quark narrow width approximation is assumed. Thus, we can have the spin-analyzing power $P^{{\rm VLQs}}_{l}$ for the decay chain of VLQs,
\begin{align}
P^{{\rm VLQs}}_{l}=P^{{\rm VLQs}}_{t} \cdot P^{t}_{l}
\end{align}
Since the direction of the charged lepton momentum in the decay of $t\rightarrow b\,l^{+}\nu$ is totally correlated with top quark polarization at leading order\,\cite{Jezabek:1994zv}, $P^{t}_{l}=1$ is used in our calculations.

\subsection{T singlet}
In the $T$ singlet scenario, we focus on the decay chains $T\rightarrow Zt\rightarrow Z(b\,l^{+}\nu)$ and $T\rightarrow Ht\rightarrow H(b\,l^{+}\nu)$. We first calculate the normalized differential decay width of $T$ decay into $Zt$
\begin{align}
\frac{1}{\Gamma}\frac{d\Gamma}{d\cos\theta_{t}}=\frac{1}{2}\left[1+\frac{|\vec{p}|(m^{2}_{Z}+2E_{T}E_{Z}-2|\vec{p}|^{2})}{2E_{T}E^{2}_{Z}-2E_{Z}|\vec{p}|^{2}+m^{2}_{Z}E_{T}}\cos\theta_{t}\right],
\end{align}
where $\theta_{t}$ is the angle between the momentum of top quark and the spin of $T$ quark in the center-of-mass system. $\vec{p}$ is the momentum of a final-state particle and $E_{T/Z}$ the energy of $T$ quark/Z in the C.M.S. which can be expressed simply using the parameter $\lambda$
\begin{align}
|\vec{p}|={}&\frac{\lambda^{1/2}(m^{2}_{T},m^{2}_{t},m^{2}_{Z})}{2m_{t}}\,,\\
E_{T}={}&\frac{m^{2}_{T}+m^{2}_{t}-m^{2}_{Z}}{2m_{t}}\,,\\
E_{Z}={}&\frac{m^{2}_{T}-m^{2}_{t}-m^{2}_{Z}}{2m_{t}}\,,\\
\lambda(a,b,c)\equiv{}& a^{2}+b^{2}+c^{2}-2ab-2bc-2ac\,.
\end{align}
Then, we can have the spin-analyzing power $P_{l}$ for the decay chain $T\rightarrow Zt\rightarrow Z(b\,l^{+}\nu)$,
\begin{align}
P^{T\,\textrm{singlet}}_{l,\,T\rightarrow Zt}=\frac{|\vec{p}|(m^{2}_{Z}+2E_{T}E_{Z}-2|\vec{p}|^{2})}{2E_{T}E^{2}_{Z}-2E_{Z}|\vec{p}|^{2}+m^{2}_{Z}E_{T}}\,.\label{pl-T-Zt}
\end{align}

For the decay chain $T\rightarrow Ht\rightarrow H(b\,l^{+}\nu)$ a similar calculation is performed and we have the spin-analyzing power of the charged lepton
\begin{align}
P^{T\,\textrm{singlet}}_{l,\,T\rightarrow Ht}=\frac{|\vec{p}|(m^{2}_{T}-m^{2}_{t})}{2m_{t}m^{2}_{T}+E_{T}(m^{2}_{T}+m^{2}_{t})}\,,\label{pl-T-Ht}
\end{align}
where $\vec{p}$ in this equation is the momentum of the final-state particle in the decay $T\rightarrow Ht$ in the C.M.S. $E_{T}$ is the energy of $T$ quark in this process.

\subsection{B singlet}
In the $B$ singlet scenario, we calculate the spin-analyzing power of the final charged lepton for the process $B\rightarrow W^{-}t\rightarrow W^{-}(b\,l^{+}\nu)$. A similar result is obtained with replacement of $m_{W}\rightarrow m_{Z}$ and $E_{T}\rightarrow E_{B}$ in \eqref{pl-T-Zt},
\begin{align}
P^{B\,\textrm{singlet}}_{l,\,B\rightarrow Wt}=\frac{|\vec{p}|(m^{2}_{W}+2E_{B}E_{W}-2|\vec{p}|^{2})}{2E_{B}E^{2}_{W}-2E_{W}|\vec{p}|^{2}+m^{2}_{W}E_{B}}\,,\label{pl-B-Wt}
\end{align}
with $\vec{p}$ defined similarly as above.

\subsection{(TB) doublet}
In the $(TB)$ doublet scenario, the spin-analyzing power is calculated for all of the above three processes $T\rightarrow Z/H t\rightarrow Z/H (b\,l^{+}\nu)$ and $B\rightarrow W^{-}t\rightarrow W^{-}(b\,l^{+}\nu)$. As can be seen from the couplings we give in section II, the left-handed couplings in the $T/B$ singlet scenarios are turned to be right-handed in the $(TB)$ doublet scenario. Following the similar calculation in the singlet scenarios, we have the spin-analyzing power in the $(TB)$ doublet scenario
\begin{align}
P^{TB\,\textrm{doublet}}_{l,\,T\rightarrow Zt}={}&\frac{|\vec{p}|(2|\vec{p}|^{2}-m^{2}_{Z}-2E_{T}E_{Z})}{2E_{T}E^{2}_{Z}-2E_{Z}|\vec{p}|^{2}+m^{2}_{Z}E_{T}}\,\quad\textrm{for}\,\,T\rightarrow Zt\rightarrow Z(b\,l^{+}\nu)\,,\label{pl-TB-TZt}\\
P^{TB\,\textrm{doublet}}_{l,\,T\rightarrow Ht}={}&\frac{|\vec{p}|(m^{2}_{t}-m^{2}_{T})}{2m_{t}m^{2}_{T}+E_{T}(m^{2}_{T}+m^{2}_{t})}\,\quad\textrm{for}\,\,T\rightarrow Ht\rightarrow H(b\,l^{+}\nu)\,,\label{pl-TB-THt}\\
P^{TB\,\textrm{doublet}}_{l,\,B\rightarrow Wt}={}&\frac{|\vec{p}|(2|\vec{p}|^{2}-m^{2}_{W}-2E_{B}E_{W})}{2E_{B}E^{2}_{W}-2E_{W}|\vec{p}|^{2}+m^{2}_{W}E_{B}}\,\quad\textrm{for}\,\,B\rightarrow Wt\rightarrow W(b\,l^{+}\nu)\,.\label{pl-TB-BWt}
\end{align}
From \eqref{pl-T-Zt}-\eqref{pl-TB-BWt} we can see that spin-analyzing power of the final charged lepton in the doublet scenarios is turned opposite from one in the singlet scenario for the corresponding processes, since the VLQ-top couplings in the singlet and doublet scenarios have the opposite handedness.

\subsection{(XT) doublet}
\begin{figure}
\includegraphics[scale=1.2]{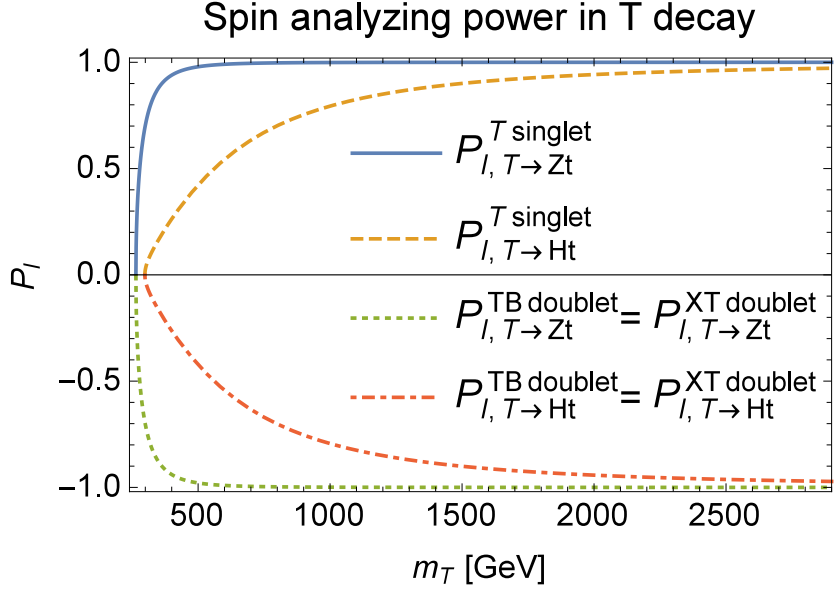}
\caption{Spin-analyzing power as a function of $m_{T}$ in T decay.}
\label{fig:all-T}
\end{figure}

\begin{figure}
\includegraphics[scale=1.2]{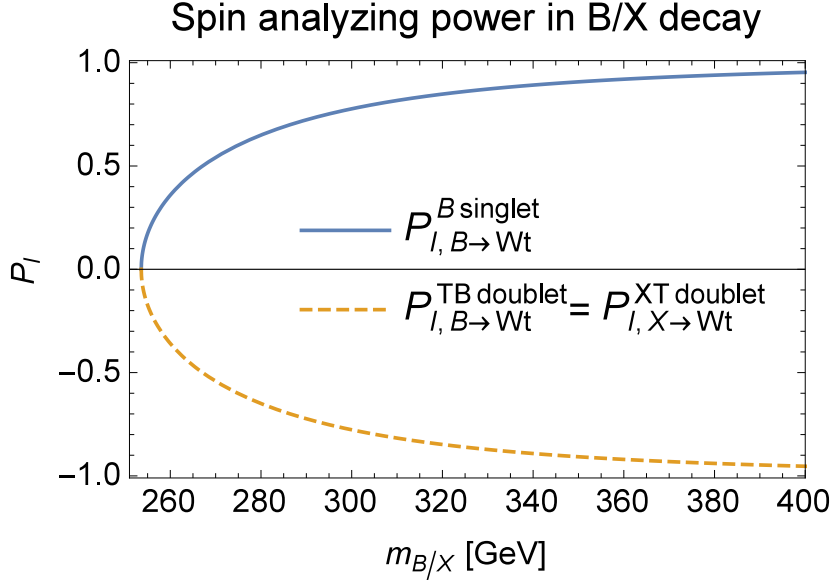}
\caption{Spin-analyzing power as a function of $m_{B/X}$ in B/X decay.}
\label{fig:all-BX}
\end{figure}

In the $(XT)$ doublet scenario, calculation is performed for three processes $X\rightarrow W^{+}t\rightarrow W^{+}(b\,l^{+}\nu)$, $T\rightarrow Zt\rightarrow Z(b\,l^{+}\nu)$ and $T\rightarrow Ht\rightarrow H(b\,l^{+}\nu)$. With replacement of $m_{B}\rightarrow m_{X}$ in \eqref{pl-TB-TZt}-\eqref{pl-TB-BWt} for the $(TB)$ doublet scenario, we arrive at the results for the $(XT)$ doublet scenario
\begin{align}
P^{XT\,\textrm{doublet}}_{l,\,T\rightarrow Zt}={}&\frac{|\vec{p}|(2|\vec{p}|^{2}-m^{2}_{Z}-2E_{T}E_{Z})}{2E_{T}E^{2}_{Z}-2E_{Z}|\vec{p}|^{2}+m^{2}_{Z}E_{T}}\,\quad\textrm{for}\,\,T\rightarrow Zt\rightarrow Z(b\,l^{+}\nu)\,,\label{pl-XT-TZt}\\
P^{XT\,\textrm{doublet}}_{l,\,T\rightarrow Ht}={}&\frac{|\vec{p}|(m^{2}_{t}-m^{2}_{T})}{2m_{t}m^{2}_{T}+E_{T}(m^{2}_{T}+m^{2}_{t})}\,\quad\textrm{for}\,\,T\rightarrow Ht\rightarrow H(b\,l^{+}\nu)\,,\label{pl-XT-THt}\\
P^{XT\,\textrm{doublet}}_{l,\,X\rightarrow Wt}={}&\frac{|\vec{p}|(2|\vec{p}|^{2}-m^{2}_{W}-2E_{X}E_{W})}{2E_{X}E^{2}_{W}-2E_{W}|\vec{p}|^{2}+m^{2}_{W}E_{X}}\,\quad\textrm{for}\,\,X\rightarrow Wt\rightarrow W(b\,l^{+}\nu)\,.\label{pl-XT-XWt}
\end{align}
As for the $(BY)$ doublet scenario there are no allowed decay processes from VLQ to top quark, so this scenario is not discussed here.
To have an intuitive understanding on the polarization effect of top quark from VLQ decay, we plot the calculated spin-analyzing powers as functions of the VLQ masses. In ~\figurename\ref{fig:all-T} we show $P_{l}$'s for $T$ decay in different scenarios as a function of the mass of $T$ quark. And in ~\figurename\ref{fig:all-BX} we show $P_{l}$'s for $B/X$ decay in different scenarios as a function of the mass of $B/X$ quark.
Actually due to the closeness of $m_{Z}$ and $m_{W}$, the distribution of spin-analyzing power corresponding to $T\rightarrow Zt\rightarrow Z(b\,l^{+}\nu)$ in $T$ singlet scenario and the one corresponding to $B\rightarrow W^{-}t\rightarrow W^{-}(b\,l^{+}\nu)$ in $B$ singlet scenario are pretty close to each other, which may not be obvious if we compare the above two figures. And similarly the curve corresponding to $T\rightarrow Zt\rightarrow Z(b\,l^{+}\nu)$ in $(TB)/(XT)$ doublet is quite close to the one corresponding to $B\rightarrow Wt\rightarrow W^{-}(b\,l^{+}\nu)$ in the same scenarios.

From Lagrangians given in section II.A, one can see that top quark from VLQ decay in the singlet scenarios is always left-handed. While in the doublet scenarios, top quark form VLQ decay is always right-handed due to the right-handed couplings given in section II.B. In $T\rightarrow Zt$ or $B/X\rightarrow W^{\mp}t$ decays, the top quark's spin state is determined by the helicity of vector bosons: for a longitudinally polarized $Z/W$, the directions of top spin and $T/B/X$ spin are parallel; while for a left-handed $Z/W$, top spin and $T/B/X$ spin directions are anti-parallel. In the $T\rightarrow Ht$ decay, top spin can be either parallel or anti-parallel to the $T$ spin since their coupling structure is a mixture of left- and right-handedness. In all of the cases, as the mass of the VLQ increases, curves of spin-analyzing power approach to $1$ (or $-1$). And this is what one expects since the more massive the VLQ is, the more the top is boosted and the top decay, as mentioned above, has the maximal spin-analyzing power ($\sim 1$) for the charged lepton. It should also be noted that, spin-analyzing power for the charged lepton from $T\rightarrow Ht\rightarrow H(b\,l^{+}\nu)$ grows much slower than the one from gauge boson mediated decay in both singlet and doublet scenarios, which is due to the fact that the VLQ-gauge couplings are purely chiral whereas the VLQ-Higgs couplings contain both left- and right-handed components. In summary, in case VLQ decay processes are probed at colliders, the polarization effects of the top quark can serve to determine its coupling structure with the VLQ.

\section{conclusion}

In this paper, we calculate the spin-analyzing power for the charged lepton from top quark in the VLQ decay in singlet and doublet VLQ scenarios. We find that the top polarization effect is helpful to differentiate various VLQ couplings with the SM particles. The spin-analyzing power for the final charged lepton is positive for the singlet VLQ scenarios, while for the doublet VLQ scenarios it is negative. Calculation also shows that spin-analyzing power for charged lepton from Higgs mediated VLQ decay grows slower than the one from gauge boson mediated VLQ decay, as a result of the difference between the chiral structures of their interactions.

\section*{Acknowledgement}
This work is supported by the National Natural Science Foundation of China (NNSFC) under grants No.\,11847208 and No.\,11705093, as well as Jiangsu Specially Appointed Professor Program.


\begin{thebibliography}{99}

\bibitem{Perelstein:2003wd}
  M.~Perelstein, M.~E.~Peskin and A.~Pierce,
  Phys.\ Rev.\ D {\bf 69}, 075002 (2004)

\bibitem{Contino:2006qr}
  R.~Contino, L.~Da Rold and A.~Pomarol,
  Phys.\ Rev.\ D {\bf 75}, 055014 (2007)

\bibitem{Matsedonskyi:2012ym}
  O.~Matsedonskyi, G.~Panico and A.~Wulzer,
  JHEP {\bf 1301}, 164 (2013)

\bibitem{Moroi:1992zk}
  T.~Moroi and Y.~Okada,
  Phys.\ Lett.\ B {\bf 295}, 73 (1992).

\bibitem{Babu:2008ge}
  K.~S.~Babu, I.~Gogoladze, M.~U.~Rehman and Q.~Shafi,
  Phys.\ Rev.\ D {\bf 78}, 055017 (2008)

\bibitem{Martin:2012dg}
  S.~P.~Martin and J.~D.~Wells,
  Phys.\ Rev.\ D {\bf 86}, 035017 (2012)

\bibitem{Cheng:2005as}
  H.~C.~Cheng, I.~Low and L.~T.~Wang,
  Phys.\ Rev.\ D {\bf 74}, 055001 (2006)

\bibitem{Han:2003wu}
  T.~Han, H.~E.~Logan, B.~McElrath and L.~T.~Wang,
  Phys.\ Rev.\ D {\bf 67}, 095004 (2003)

\bibitem{Atre:2011ae}
  A.~Atre, G.~Azuelos, M.~Carena, T.~Han, E.~Ozcan, J.~Santiago and G.~Unel,
  JHEP {\bf 1108}, 080 (2011)

\bibitem{Atre:2008iu}
  A.~Atre, M.~Carena, T.~Han and J.~Santiago,
  Phys.\ Rev.\ D {\bf 79}, 054018 (2009)

\bibitem{Cao:2006wk}
  Q.~H.~Cao, C.~S.~Li and C.-P.~Yuan,
  Phys.\ Lett.\ B {\bf 668}, 24 (2008)

\bibitem{Cao:2007ea}
  Q.~H.~Cao, J.~Wudka and C.-P.~Yuan,
  Phys.\ Lett.\ B {\bf 658}, 50 (2007)

\bibitem{Han:2016bus}
  X.~F.~Han, L.~Wang, L.~Wu, J.~M.~Yang and M.~Zhang,
  Phys.\ Lett.\ B {\bf 756}, 309 (2016)

\bibitem{Liu:2015kmo}
  N.~Liu, L.~Wu, B.~Yang and M.~Zhang,
  Phys.\ Lett.\ B {\bf 753}, 664 (2016)

\bibitem{Han:2014qia}
  C.~Han, A.~Kobakhidze, N.~Liu, L.~Wu and B.~Yang,
  Nucl.\ Phys.\ B {\bf 890}, 388 (2014)

\bibitem{Kearney:2013oia}
  J.~Kearney, A.~Pierce and J.~Thaler,
  JHEP {\bf 1308}, 130 (2013)

\bibitem{Kearney:2013cca}
  J.~Kearney, A.~Pierce and J.~Thaler,
  JHEP {\bf 1310}, 230 (2013)

\bibitem{Kats:2017ojr}
  Y.~Kats, M.~McCullough, G.~Perez, Y.~Soreq and J.~Thaler,
  JHEP {\bf 1706}, 126 (2017)
  
\bibitem{Barducci:2017xtw}
  D.~Barducci and L.~Panizzi,
  JHEP {\bf 1712}, 057 (2017)

\bibitem{delAguila:2000aa}
  F.~del Aguila, M.~Perez-Victoria and J.~Santiago,
  Phys.\ Lett.\ B {\bf 492}, 98 (2000)

\bibitem{delAguila:2000rc}
  F.~del Aguila, M.~Perez-Victoria and J.~Santiago,
  JHEP {\bf 0009}, 011 (2000)

\bibitem{AguilarSaavedra:2009es}
  J.~A.~Aguilar-Saavedra,
  JHEP {\bf 0911}, 030 (2009)

\bibitem{delAguila:1982fs}
  F.~del Aguila and M.~J.~Bowick,
  Nucl.\ Phys.\ B {\bf 224}, 107 (1983).

\bibitem{Branco:1986my}
  G.~C.~Branco and L.~Lavoura,
  Nucl.\ Phys.\ B {\bf 278}, 738 (1986).

\bibitem{Djouadi:2012ae}
  A.~Djouadi and A.~Lenz,
  Phys.\ Lett.\ B {\bf 715}, 310 (2012)

\bibitem{Eberhardt:2012gv}
  O.~Eberhardt, G.~Herbert, H.~Lacker, A.~Lenz, A.~Menzel, U.~Nierste and M.~Wiebusch,
  Phys.\ Rev.\ Lett.\  {\bf 109}, 241802 (2012)

\bibitem{Sirunyan:2017pks}
  A.~M.~Sirunyan {\it et al.} [CMS Collaboration],
  Phys.\ Lett.\ B {\bf 779}, 82 (2018)

\bibitem{Tanabashi:2018oca}
  M.~Tanabashi {\it et al.} [Particle Data Group],
  Phys.\ Rev.\ D {\bf 98}, no. 3, 030001 (2018).

\bibitem{Aad:2015voa}
  G.~Aad {\it et al.} [ATLAS Collaboration],
  JHEP {\bf 1602}, 110 (2016)

\bibitem{Sirunyan:2017jin}
  A.~M.~Sirunyan {\it et al.} [CMS Collaboration],
  JHEP {\bf 1708}, 073 (2017)

\bibitem{V.:2016wba}
  V.~Arunprasath, R.~M.~Godbole and R.~K.~Singh,
  Phys.\ Rev.\ D {\bf 95}, no. 7, 076012 (2017)

\bibitem{Belanger:2012tm}
  G.~Belanger, R.~M.~Godbole, L.~Hartgring and I.~Niessen,
  JHEP {\bf 1305}, 167 (2013)

\bibitem{Choudhury:2010cd}
  D.~Choudhury, R.~M.~Godbole, S.~D.~Rindani and P.~Saha,
  Phys.\ Rev.\ D {\bf 84}, 014023 (2011)

\bibitem{Godbole:2006tq}
  R.~M.~Godbole, S.~D.~Rindani and R.~K.~Singh,
  JHEP {\bf 0612}, 021 (2006)

\bibitem{Han:2008gy}
  T.~Han, R.~Mahbubani, D.~G.~E.~Walker and L.~T.~Wang,
  JHEP {\bf 0905}, 117 (2009)

\bibitem{Barger:2006hm}
  V.~Barger, T.~Han and D.~G.~E.~Walker,
  Phys.\ Rev.\ Lett.\  {\bf 100}, 031801 (2008)

\bibitem{Arai:2009cp}
  M.~Arai, N.~Okada and K.~Smolek,
  Phys.\ Rev.\ D {\bf 79}, 074019 (2009)

\bibitem{Arai:2007ts}
  M.~Arai, N.~Okada, K.~Smolek and V.~Simak,
  Phys.\ Rev.\ D {\bf 75}, 095008 (2007)

\bibitem{Liu:2010zze}
  J.~Y.~Liu, Z.~G.~Si and C.~X.~Yue,
  Phys.\ Rev.\ D {\bf 81}, 015011 (2010).

\bibitem{Gopalakrishna:2010xm}
  S.~Gopalakrishna, T.~Han, I.~Lewis, Z.~g.~Si and Y.~F.~Zhou,
  Phys.\ Rev.\ D {\bf 82}, 115020 (2010)

\bibitem{Bernreuther:2006pd}
  W.~Bernreuther, M.~Fuecker and Z.~G.~Si,
  PoS TOP {\bf 2006}, 018 (2006).

\bibitem{Bernreuther:2010ny}
  W.~Bernreuther and Z.~G.~Si,
  Nucl.\ Phys.\ B {\bf 837}, 90 (2010)

\bibitem{Bernreuther:2013aga}
  W.~Bernreuther and Z.~G.~Si,
  Phys.\ Lett.\ B {\bf 725}, 115 (2013)
  Erratum: [Phys.\ Lett.\ B {\bf 744}, 413 (2015)]

\bibitem{Bernreuther:2015yna}
  W.~Bernreuther, D.~Heisler and Z.~G.~Si,
  JHEP {\bf 1512}, 026 (2015)

\bibitem{Bernreuther:2017yhg}
  W.~Bernreuther, P.~Galler, Z.~G.~Si and P.~Uwer,
  Phys.\ Rev.\ D {\bf 95}, no. 9, 095012 (2017)

\bibitem{He:2007tt}
  X.~G.~He, T.~Li, X.~Q.~Li and H.~C.~Tsai,
  Mod.\ Phys.\ Lett.\ A {\bf 22}, 2121 (2007)

\bibitem{Wu:2018xiz}
  L.~Wu and H.~Zhou,
  arXiv:1811.08573 [hep-ph].

\bibitem{Cao:2015doa}
  Q.~H.~Cao, B.~Yan, J.~H.~Yu and C.~Zhang,
  Chin.\ Phys.\ C {\bf 41}, no. 6, 063101 (2017)

\bibitem{Jezabek:1994zv}
  M.~Jezabek and J.~H.~Kuhn,
  Phys.\ Lett.\ B {\bf 329}, 317 (1994)

\end{thebibliography}
\end{document}